\documentclass [12pt,twoside]{article}           
\usepackage[left,modulo]{lineno}
\usepackage{setspace}

\countdef\arxiv=201

\ifnum\arxiv=0 \input cpreb.tex \fi

\ifnum\arxiv=1

\usepackage{epsfig,times,lscape}
\usepackage[usenames]{color}

    \ifnum\count102=1

    \fi

    \ifnum\count102=2
\fi

\newcommand{\nc}{\newcommand}

     \ifnum\count101=1
\nc{\qI}[1]{\section{{#1}}}
\nc{\qA}[1]{\subsection{{#1}}}
\nc{\qun}[1]{\subsubsection{{#1}}}
\nc{\qa}[1]{\paragraph{{#1}}}

\def\qpar{\vskip 2mm plus 0.2mm minus 0.2mm}
\def\qL{\hfill \break}
     \fi 

      \ifnum\count101=2
 \nc{\qI}[1]{\parindent=0mm \vskip 8mm 
{\centerline{\LARGE \color{red}#1}}\vskip 3mm}
\nc{\qA}[1]{\vskip 2.5mm \noindent 
{{\bf\large\color{blue}  #1}} \vskip 1mm \parindent=0mm}
 \nc{\qun}[1]{\vskip 1mm \noindent {\sl #1 }\quad }

\def\qL{\hfill \break}
\def\qpar{\vskip 2mm plus 0.2mm minus 0.2mm}

      \fi

\def\qth{\vrule height 12pt depth 0pt width 0pt}
\def\qtb{\vrule height 0pt depth 5pt width 0pt}

\nc{\qfoot}[1]{\footnote{{#1}}}

      \ifnum\count101=1
\def\qbu{\hfill \par \hskip 6mm $ \bullet $ \hskip 2mm}

      \fi
      \ifnum\count101=2
\def\qbu{\hfill \par \hskip 4mm $ \bullet $ \hskip 2mm}

      \fi

\def\qparr{ \vskip 1.0mm plus 0.2mm minus 0.2mm \hangindent=10mm
\hangafter=1}

     \ifnum\count101=1 
 \def\qdec#1{\parindent=0mm\par {\leftskip=2cm {#1} \par}}
     \fi
     \ifnum\count101=2
  \def\qdec#1{\parindent=0mm \par {\leftskip=1cm {#1} \par}}
  
  \def\qcitb#1{\noindent \hbox to 102mm{\hfill \small #1} \vskip 1mm}
      \fi

 \def\qpages#1{\count102=0{\loop\advance\count102 by 1
 \null \vfill\eject \ifnum\count102<#1 \repeat}}

\def\qth{\vrule height 12pt depth 0pt width 0pt}
\def\qtb{\vrule height 0pt depth 5pt width 0pt}

\def\qv{\vskip 0.1mm plus 0.05mm minus 0.05mm}
\def\qhu{\hskip 0.6mm}
\def\qhv{\hskip 3mm}

\def\qhw{\hskip 1.5mm}
\def\qleg#1#2#3{\noindent {\bf \small #1\qhw}{\small #2\qhw}{\it \small #3}\qv }
\fi

\begin{document}
\thispagestyle{empty}



\markboth{{\sl \hfill  \hfill \protect\phantom{3}}}
        {{\protect\phantom{3}\sl \hfill  \hfill}}

\color{yellow} 
\hrule height 10mm depth 10mm width 170mm 
\color{black}

\vskip -12mm 

\centerline{\bf \large Excess-tuberculosis-mortality in young women:
high accuracy exploration}
\vskip 15mm

\centerline{\large 
Sylvan Berrut$ ^1 $,
Peter Richmond$ ^2 $ and Bertrand M. Roehner$ ^3 $
}

\vskip 10mm
\large

%
{\bf \color{red} SUMMARY}\qL
{\bf \color{blue} Background:} \quad
In a general way at all ages and for almost all
diseases, male death rates are higher than female death rates.\qL
{\bf \color{blue} Findings:}\quad 
Here we report a case in which the opposite holds, namely
for tuberculosis (TB) mortality 
between the ages of 5 and 25, female death rates are about
two times higher than male rates.
What makes this observation
of interest is that it occurs in all countries for which
data are available (e.g. Britain, Switzerland and United States),
and in all years from the end of the 19th century up to
the time in the 1960s when TB became a very rare
disease in all developed countries. The fact that this regularity
holds despite a drastic reduction in the number of deaths is
also noteworthy.\qL
{\bf \color{blue} Practical usefulness:}\quad 
So far, the reason
of this anomaly remains an open question but the
effect is so accurate that it can be used for probing
the reliability of mortality records. This will
be explained in the case of developing countries.
For instance, it turns out that
in South African TB death data as published
(and revised) 
by the ``World Health Organization'', 
female deaths were certainly under-estimated by a factor of two.

\vskip 2mm
\centerline{\it \small Version of 31 January 2018}
\vskip 2mm

{\small Key-words: death rate, death ratio, male, female, 
tuberculosis}

\vskip 5mm

{\normalsize
1: Swiss Federal Office of Statistics, Neuch\^atel, Switzerland.
Email: Sylvie.Berrut@bfs.admin.ch \qL
2: School of Physics, Trinity College Dublin, Ireland.
Email: peter\_richmond@ymail.com \qL
3: Institute for Theoretical and High Energy Physics (LPTHE),
University Pierre and Marie Curie, Paris, France. 
CNRS (Centre National de la Recherche Scientifique),
UMR 7589 (Unit\'e Mixte de Recherche).
Email: roehner@lpthe.jussieu.fr
}

\vfill\eject

\qI{Introduction}

Because
this question can serve to illustrate the methodological
discrepancy between the approach of physics%
\qfoot{Above all the approach of physics means
accurate  well targeted
measurements along with their comparative analysis.}
on the one hand
and of biology on the other hand, we believe its
significance extends beyond the specific problem investigated
here. 
\qpar

After a short presentation of the excess-female mortality
effect, in the two following subsections
we explain why the answers proposed by
epidemiologists and biologists appear unsatisfactory.
Actually, 
the reason of this failure is quite simple; it is the inability
or unwillingness to adopt a comparative perspective.

\qA{Excess-female mortality in tuberculosis}

In human populations females have a lower mortality
than males at any age and for almost all causes of death.
However there are a few exceptions and it is therefore
natural that they attracted the attention of epidemiologists.
In the present paper we concentrate our attention on the
fact that between the age of 5 and 25 the tuberculosis
death rate of females is about 1.5 to 2.0 times higher than for males.
It is true that there are a number of other infectious
diseases (see below) which show a similar effect but
almost all such diseases are childhood diseases (e.g. measles) 
which means 
that most of the cases occur prior to the age of 5.
For instance, in 1901 in the UK, 94\% of the measles deaths
occurred between the ages of 0 and 5. In the $ (15-44) $ age
interval, the only one in which there is a substantial excess
female mortality, there are only 42 deaths (0.5\% of the annual
number).
In other words, for all those diseases except tuberculosis,
the excess-female mortality effect
concerns a very small number of patients. On the contrary,
for tuberculosis the age-specific mortality increases from the 
age of 5 to the age of 25 which means that, especially
in the developing world, hundreds of thousands of patients
are concerned by the female excess mortality effect.

\qA{Poor diet as a first suggested explanation}

Several papers published in the past three decades from
1990 to 2017 (Anderson 1990, Hinde 2011, Janssens 2017)
propose an explanation based on an inappropriate diet.
The thesis can be summarized as follows (Hinde 2011, p.9).
\qdec{
``A widely held account is that a lack of bargaining power in the home
associated with a
shortage of paid work for women led to women having a
much poorer diet than men, which lowered their resistance to
infections. In 1990
Michael Anderson argued that this was the underlying reason for the
relatively high
female mortality compared to that of males observed 
[in 19th century Britain] in poor agricultural areas and
regions dominated by heavy industry.''}
\qpar

A similar thesis was presented independently 
for the Netherlands by Ang\'elique Janssens:
\qdec{``We have been able to ascertain that a considerable part of
maternal mortality in the period 1875-1900 can be attributed
respiratory TB for which adequate nutritional intakes are 
highly relevant.''}
\qpar

What should one think of this explanation?\qL
A piece of evidence that would provide 
a solid basis for this explanation would consist in
statistics about the respective food intakes of boys and girls.
Needless to say, no data of that kind are available
at country level.
Thus, one must rely on the belief that it made sense for
families to favor their sons at the expense of their daughters.
\qpar
However, the main problem with this explanation is the fact that the
excess-female mortality is not limited to the 19th century.
In fact, as will be shown below in Fig. 1, it extends well into
the 20th century including to areas like California or New York
State which are not ``poor agricultural areas''. By extending their
analysis to the 20th century, the aforementioned authors
would have been able to identify this difficulty.
\qpar

In short, we do not want to say that the diet effect played
no role whatsoever, but it is certainly not the main
explanation for this effect.

\qA{An explanation based on the role of sex hormones}

In a sense the excess-female mortality
effect could be called the ``Garenne effect''
after the name of the epidemiologist Michel Garenne
who from 1991 to 1998 (Garenne et al. 1991,
Garenne 1994, Garenne et al. 1998) devoted much time
to the study of this effect. 
\qpar
The study started from an observation
made in rural Senegal in 1990 showing female excess mortality
after measles vaccination. Then, eight years later an
explanation was proposed.
\qpar

Positing a link between the hormonal and immunological
systems seems of course a natural explanation for an effect
which affects male and females differently.
\qpar

In Garenne and al. (1998) the authors say that ``there
is growing evidence that sex hormones regulate the Th1/Th2
balance''. The Th1 and  Th2 (T=Thymus where they
originate, h=helper) are two
sorts of white cells which eliminate foreign bodies,
e.g. cells infected by a virus or cancer cells. It turns out
that the female hormone progesterone promotes the 
production of Th2 cells whereas the male hormone testosterone
rather favors the Th1 cells. To close the argument one only
needs to observe that the Th2 cells seem to develop
weaker resistance to bacteria than that provided
by Th1 cells.
\qpar

Obviously, this model was designed for the specific
purpose of explaining the excess-female mortality
in an age interval during which the concentration
of progesterone in the blood is fairly high.
In order to be really convincing this model
must explain {\it other effects} than the one for
which it was designed. Unfortunately, the
authors do not give any corroborating evidence.
Once again, this denotes a lack of comparative
perspective. In physics, when a new effect has been 
identified, the first task is to define its range
of validity. For instance, surface tension explains
why some insects (e.g. water striders) can walk on water 
and it explains also the ``tears of wine'' phenomenon.
\qpar

Starting from the fact that high levels of
progesterone inhibit the fight against TB-like bacteria,
what additional observational tests can  one propose?
\qpar

First, we must find situations in which the progesterone
level is particularly high. Such situations should
be marked by an excess-female vulnerability to TB.
(i) In men or postmenopausal women the concentration of
progesterone in the serum of the blood is of the
order of 1 nanogramme per milliliter of blood serum.
(ii) For young non-pregnant females it is of the order
of 5 ng/ml (low at the beginning of the 28-day cycle
and higher at the end). (iii) Finally, during pregnancy
it is on average of the order of 50 ng/ml. 
In short, the concentration is really much higher than
in men only during pregnancy. 
\qpar

Thus, a testable prediction of the progesterone model
would be a high incidence of TB in women who have several
children in succession as compared with women
who have only one child or none at all. Naturally all
other conditions should be similar and in addition the test
should be made on the years
before the BCG vaccination became commonly used.
\qpar

The authors of Garenne et al. (1998) mention 
a fact which, at first sight, 
seems to go in the right direction. 
They say that diseases
such as rubella , influenza and tuberculosis ``are more severe
during pregnancy when the level of progesterone dramatically
increases'' but they give no evidence apart from the
well known case of rubella, which however is of a different
nature in  the sense that it is the embryo which suffers
rather than the mother.

\qA{Death rate, incidence rate, fatality rate}

 In the two previous subsections we discussed successively a 
sociological and a biological explanation, More
generally, the death rate $ \mu $ can be decomposed in the following
way:

$$ \mu={ \hbox{deaths}\over \hbox{population} },\
F={\hbox{deaths}\over \hbox{affected population} },\
I={\hbox{affected population}\over \hbox{population} } $$

$$ \hbox{Death rate }(\mu)=\hbox{Fatality rate }(F) \times 
\hbox{Incidence rate } (I) $$

The fatality rate which describes the severity of the disease
is likely to be of biological nature, whereas the incidence rate
which describes how fast the disease spreads is a mixed
factor which reflects both biological features, 
e.g. the mode of transmission
and sociological conditions, e.g. the frequency of social interactions.

\qA{Accurate measurements}

In previous studies the data
from many countries and many years (typically 1950-1989)
were lumped together. Thus, in Garenne (1994) the smallest
areas considered are whole continents: Europe, North America
(which mixes the US and Mexico), Latin America and so on.
In Garenne et al. (1998) the graphs are drawn for the whole
world.
Such  a procedure of data aggregation precluded
any serious comparative analysis.
\qpar

On the contrary, in the present paper we consider single
countries (or even subareas of countries) over time
intervals that were especially calibrated to
be the smallest intervals able to keep the statistical
fluctuations at an acceptable level.
\qpar

This procedure will allow us to make a number of preliminary
observations. Moreover, 
more precise conclusions will become possible
once incidence and fatality rate data become
available.

\qI{Anomaly of the female/male death ratio for TB: evidence}

As already said,
broadly speaking for most age groups and most diseases
the number of female deaths is markedly smaller than
the number of male deaths%
\qfoot{This observation seems to hold not only for humans but also
more generally for primates (Bronikowski et al. 2011), mammals  
and vertebrates (Clutton-Brock et al. 2007.}%
.
In the US, in the 1950s
for all causes of death and averaged over
the age interval 5-30 the death ratio male/female
is about 1.5.
\qpar

However, here
we report a case in which 
there are about
two times more female deaths than male deaths.
This effect is seen between 1880 and 1960
for tuberculosis (TB) between the ages
of 10 and 25 (Fig. 1a,b,c,d).
In the decades after 1970 the number of 
deaths due to TB became very small (except
in old age). For instance, in the US
in the time interval 1999-2015 and for
all the age groups from birth to 45 years
the annual
number of deaths due to respiratory TB averaged
only about 40. For that reason the effect
becomes impossible to test after 1960.
\qpar
The fact that this effect is seen not only in the US but
also in Switzerland and Britain
indicates that it is probably not due to a
statistical artifact. For the sake of brevity it will be
referred to as the $ f/m $ effect.
%
\begin{figure}[htb]
\centerline{\psfig{width=16cm,figure=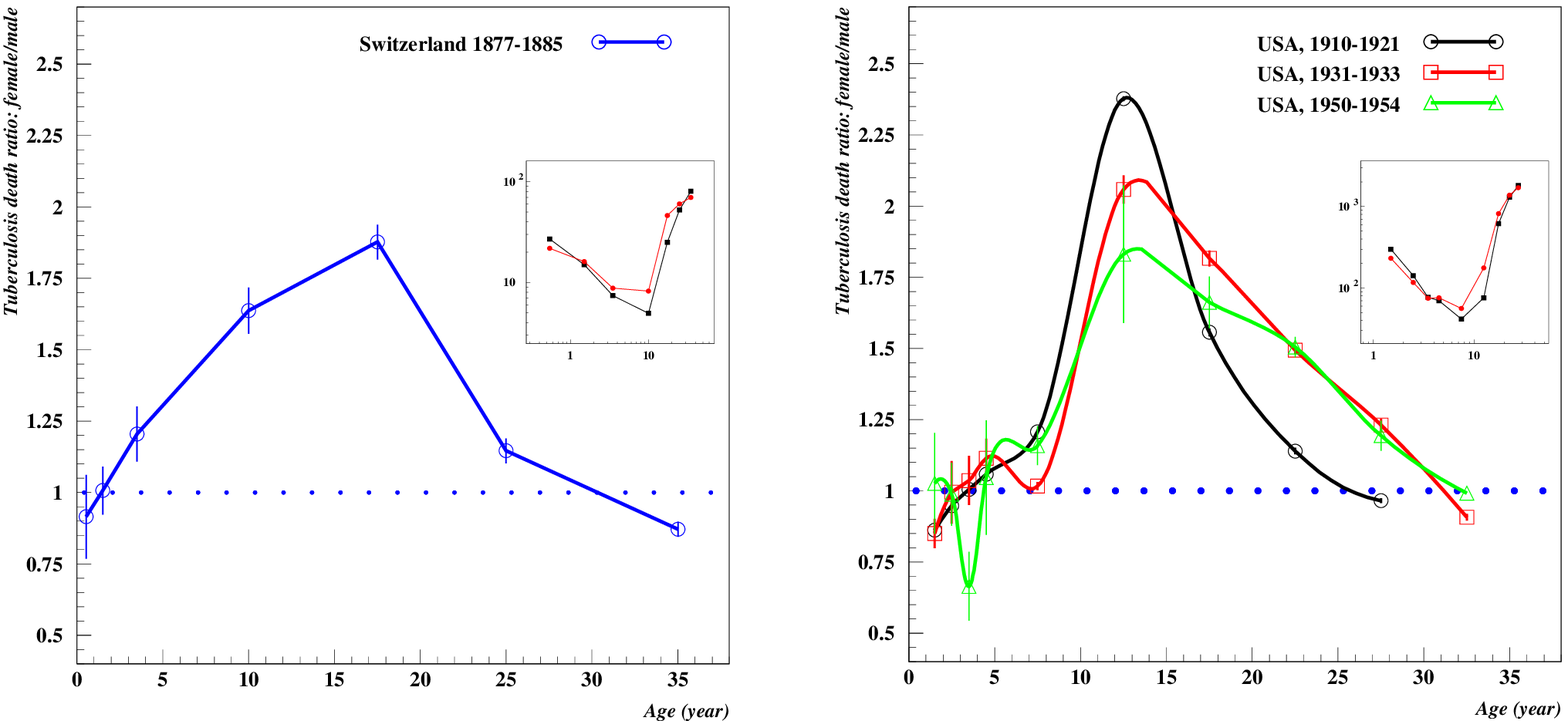}}
\qleg{Fig.\qhu 1a,b\qhv TB
death ratio female/male
by age in Switzerland and the US.}
{More precisely TB refers to
TB of the lungs. In the Swiss statistics it is the
word ``phthisis'' which is
used instead of TB.
The error bars are defined as $ \pm \sigma $ where
$ \sigma $ is the standard deviation of the average.
The insets show separately
the female (upper line in red) and male (lower
line in black) TB death rates.}
{Sources: Switzerland: Mouvement de la population de la
Suisse 1877-1885, the data are available on the website of the
``Office F\'ed\'eral de la Statistique'',
in the years after 1885 the death data by cause, age and sex were
no longer included. USA: Bureau of the Census, Mortality Statistics,
various years.}
\centerline{\psfig{width=16cm,figure=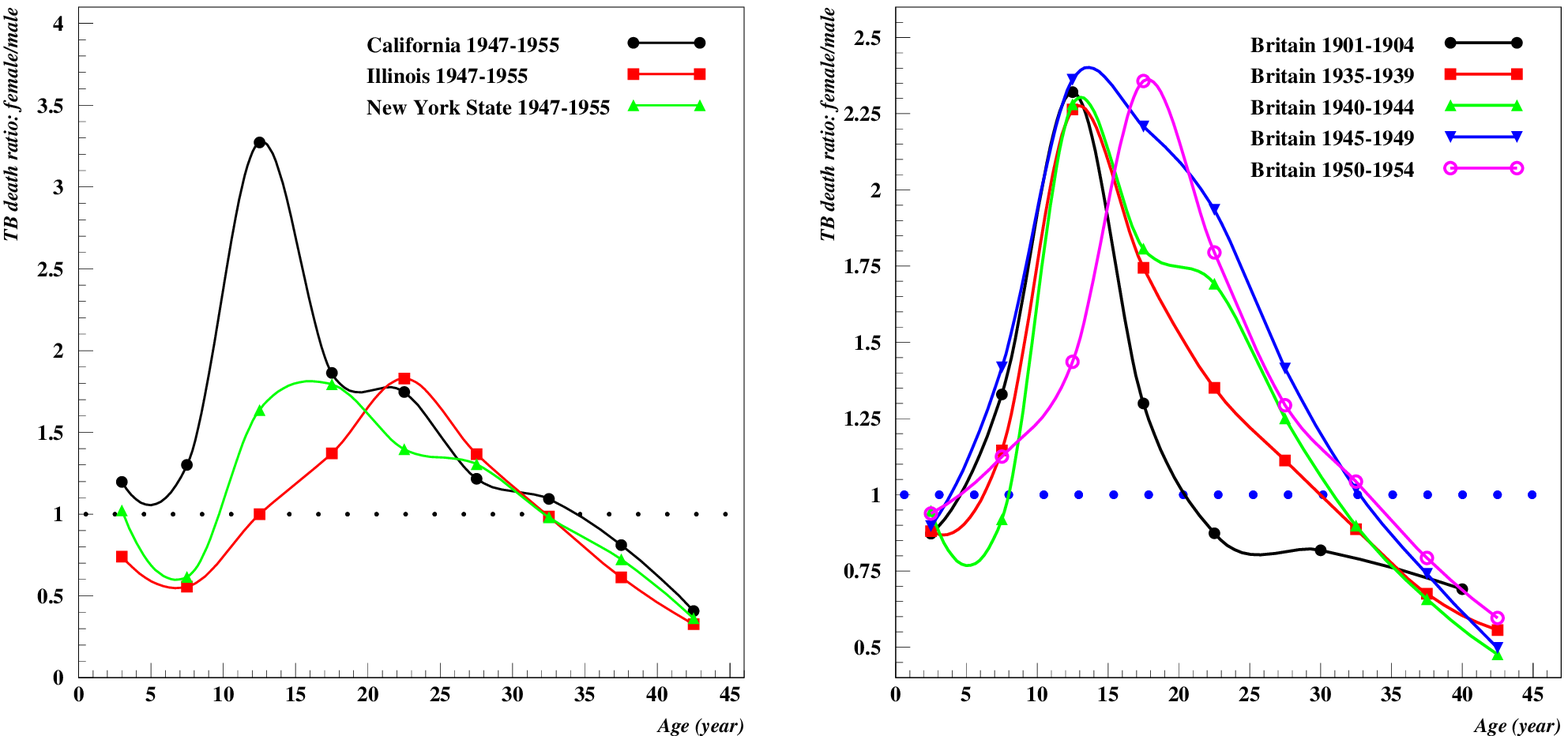}}
\qleg{Fig.\qhu 1c,d\qhv TB
death ratio female/male
by age in 3 US states (white population only)
and in Britain.}
{The error bars are rather small and
have been omitted for the sake of clarity; on average
the coefficient of variation is of the order of 4\%.}
{Sources:  Vital Statistics of the United States;
website of the British ``Office of National Statistics'', 
many thanks
to Ms. Justine Pooley for her help.}
\end{figure}

%
\begin{figure}[htb]
\centerline{\psfig{width=12cm,figure=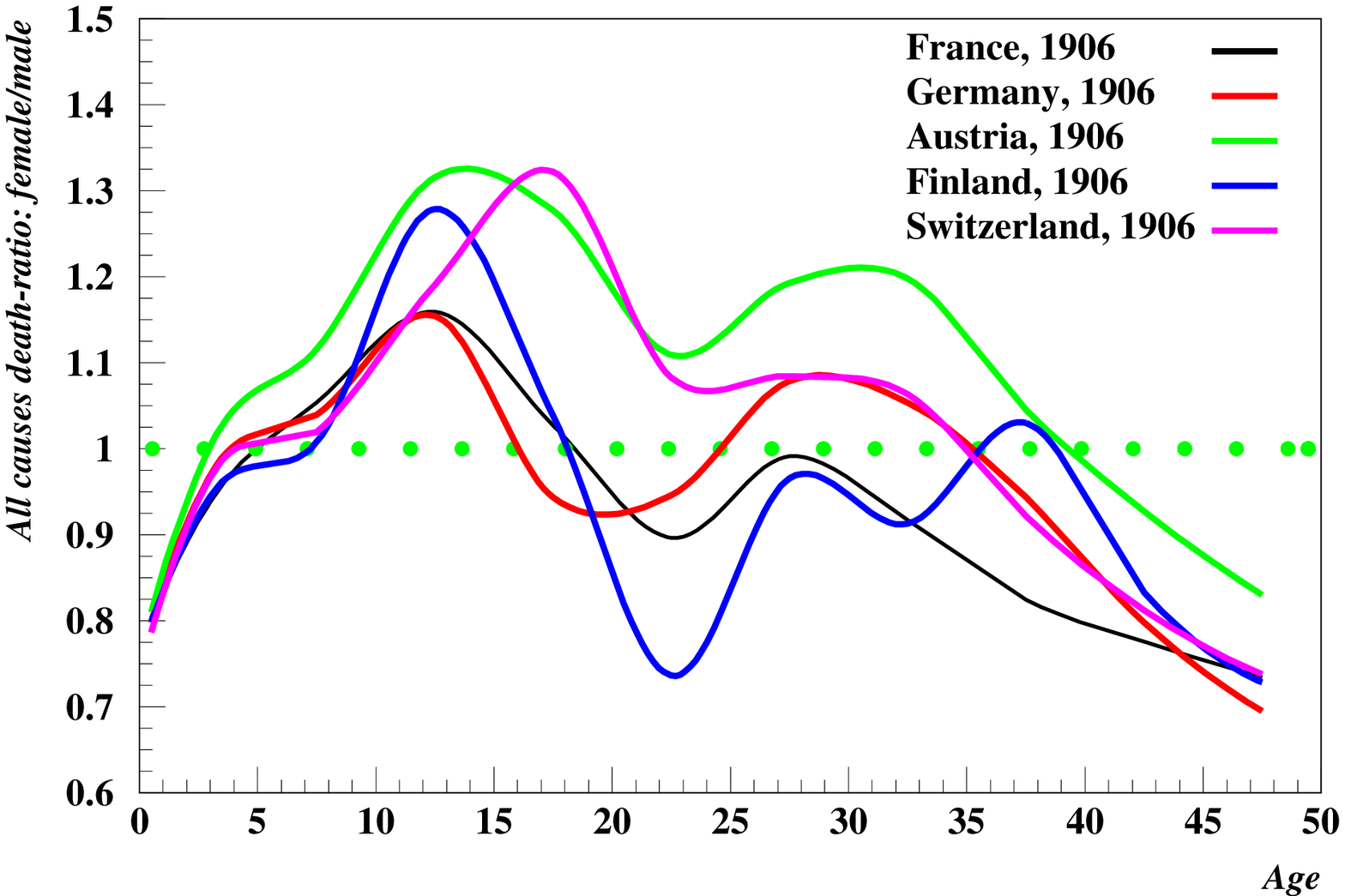}}
\qleg{Fig.\qhu 2\qhv All causes
death ratio, female/male}
{The fact that the curves follow fairly closely the TB death ratios
(although of course with a smaller amplitude)
shows the key role played by TB mortality.}
{Source: Bunle (1954).}
\end{figure}

\qpar
Fig. 2 shows the $ f/m $ death ratio for all causes of death
for several European countries%
\qfoot{In addition it can be noted that for Japan the
curve is very similar.}%
.
It can be seen that 
the age interval in which $ f/m>1 $ is basically determined
by the TB death ratio.

\qI{Discussion of the effect}

\qA{Female excess or male curtailment?}

Is this effect due to an excess of female deaths or to
an ``abnormally'' low number of male deaths? 
In an attempt to answer this question
separate male and female data are shown in Fig. 1a,b.
(male in black, female in red).
Now, what should one understand by ``normal'' curves?
The term ``normal'' is understood here as ``most common''.
The common pattern is that in a log-log plot the infant
death rate is a downward straight line until the age of
10 after which it starts fairly suddenly to go upward
(Berrut et al. 2016). The inset graphs in Fig. 1a,b
show that the female curves begin to level off already
around the age of 4. As a result they come above the curves
of the male death rates and remain higher until
the age of 25. Thus, 
it is the female rate (not the male rate)
which behaves in an unusual way. In other words, the
phenomenon is indeed an excess-female mortality. 

\qA{Exogenous or endogenous?}

As always in such situations, once the possible incidence of
a recording artifact has been excluded, the
effect can be due to exogenous or endogenous factors.
Here exogenous would mean more contacts with pathogens
or other external factors
whereas endogenous would refer to physiological factors.
\qpar

A possible exogenous factor which comes to mind is
an occupational hazard. Let us examine this possibility
more closely.\qL
The age interval over which the effect is observed
is fairly broad. However, it is the starting age which really
matters. Why? Because, once TB cases in excess have appeared
the disease will develop through its own dynamic in the sense
that it will progress in each affected individual and 
also spread by contagion to other persons.
If this argument is accepted, how should one
determine the starting age? In the graph the data points are
shown with error bars which correspond to $ \pm \sigma $.
In order to raise the confidence level from 0.68 
(which corresponds to $ \pm \sigma $ to 0.96 (which 
corresponds to $ \pm 2\sigma $
we will consider that the starting point occurs when 
the ratio becomes equal to $ 1+2\sigma $. This leads to a starting
age of 5 in Switzerland and 6 in the US. The important point
is that these ages are well before 
any industrial occupational activity. The only kind
of activity that one can think of at that age would be
domestic, for instance farm work. In most countries, the age
of 6 marks the beginning of school attendance although it is
far from clear why this should lead to an $ f/m $ effect.

\qA{Are there other diseases with more female than male deaths?}

One should remember that in the first half of the 20th century
tuberculosis was the first cause of death. This makes the
previous anomaly quite noteworthy. However one needs also to
examine whether 
there are other diseases for which female deaths outnumber male deaths.
More precisely, we wish to see if
there are other diseases which show a female/male death
ratio over 1 in the age interval 5-25?\qL
A systematic investigation of British data
for all the diseases mentioned
in the  International Classification 
of 1901 leads to the following findings.
\qbu {\color{blue} Code 60, measles.}\qL
$ 0-5:\ f/m=0.88,\ 5-15:\ f/m=1.14,\ 15-44:\ f/m=2.00 $ 
\qbu {\color{blue} Code 740, anaemia and 
leucocythaemia (nowadays rather called
leukocytosis, i.e. white cells in excess).} \qL
$ f/m \sim 3 $ in
the age interval 10-45. However, there are only few deaths.
In successive 5-year
age groups the numbers of deaths
for this code number are under 100.
\qbu {\color{blue} Code 1060, cerebral haemorrhage and embolism.}\qL
$ f/m\sim 2 $ in the age interval 1-25. 
There are less than 30 deaths in each 5-year age group.
\qbu{\color{blue}  Code 1810, burns and scalds.} \qL
This was
the most puzzling finding. In the age interval 5-20,
$ f/m\sim 3 $; then in 20-45, $ f/m\sim 1 $; finally
from 45 to 85, $ f/m $ is again about 3.
\qpar

In short, apart from the case of burns, for the age intervals
under consideration, 
the other cases are of minor importance
in terms of death numbers.

\qI{Tuberculosis death ratio in developing countries} 

In developed countries TB mortality has become close
to zero except in old age%
\qfoot{In what follows age groups of older adults will be
left aside. There are two
reasons for that. The first is of course because
the $ f/m $ effect occurs in early years. In addition, 
one should remember that because of atypical clinical symptoms
the diagnosis of tuberculosis in elderly people
is rather uncertain (Thomas et al. 2001).}%
. 
However, in many developing countries TB is still an
important disease. The question that we wish to address
is whether the $ f/m $ effect can be observed in such
countries? 

\qA{Defects of the WHO data base}

Before we can answer this question we must examine what statistical
data are available. The data of the ``World Health Organization'' 
(WHO) provide a broad coverage for almost all countries
and TB features at the top of the list of diseases.
However for the objective that we have 
in mind there are three difficulties..
\qbu The data by age are limited to only three age groups, namely:
$ I_1=0-15,\ I_2= 15-60,\ I_3=60^+ $. 
\qbu In many developing countries the quality and completeness
of the data is not good. The WHO distinguishes 3 categories
which are represented by 3 colors: ``magenta'' means very
incomplete, ``cyan'' means fairly complete, ``blue'' means
good quality data. Developed countries are blue, semi-developed
countries are cyan and almost all African countries are magenta.
For our purpose this is of course most unfortunate because it means
that for the countries where TB may be most
prevalent there are in fact no reliable data. That is why
we will focus on semi-developed countries.
\qbu 
There is another cause of uncertainty which is due to the
way the data are reported. In the table the numbers of deaths
are expressed in thousands but as
there is only one decimal
digit a number such as 0.1 could mean 0.051 or 0.149 which
means that there is an uncertainty of $ (0.149-0.051)/0.1=100\% $.
In other words, the smaller the number of deaths, the lower
its accuracy%
\qfoot{Obviously this is not a sound way of reporting
because it adds a ``reporting uncertainty'' to the ``recording
uncertainty''. The data should be reported with the same
number of digits whether the figures are small or large,
e.g. 0.36 and 36 instead of 0.4 and 36.1.}%
.
For the same reason all data for developed countries are reported
as being 0.0; this  means only that the real death numbers are less
than $ 0.05 \hbox{ thousands } = 50 $. In countries with a small
population this may still represent a sizable death rate.

\qA{Test of the $ f/m $ effect in developing countries}

Because of the limitation in the number of age-groups we cannot
test the $ f/m $ effect directly. The test must be done indirectly.
How should one proceed?
\qpar
For both $ I_1 $ and $ I_2 $ he WHO data allow us to compute
the ratio $ f/m $. They are given in Table 1b for several
countries. Then, we must compare these ratios with the same
ratios for a country (for instance the US) for which the
$ f/m $ effect is observed. It is at this point that a difficulty
arises which must be considered more closely. 

%
\begin{figure}[htb]
\centerline{\psfig{width=8cm,figure=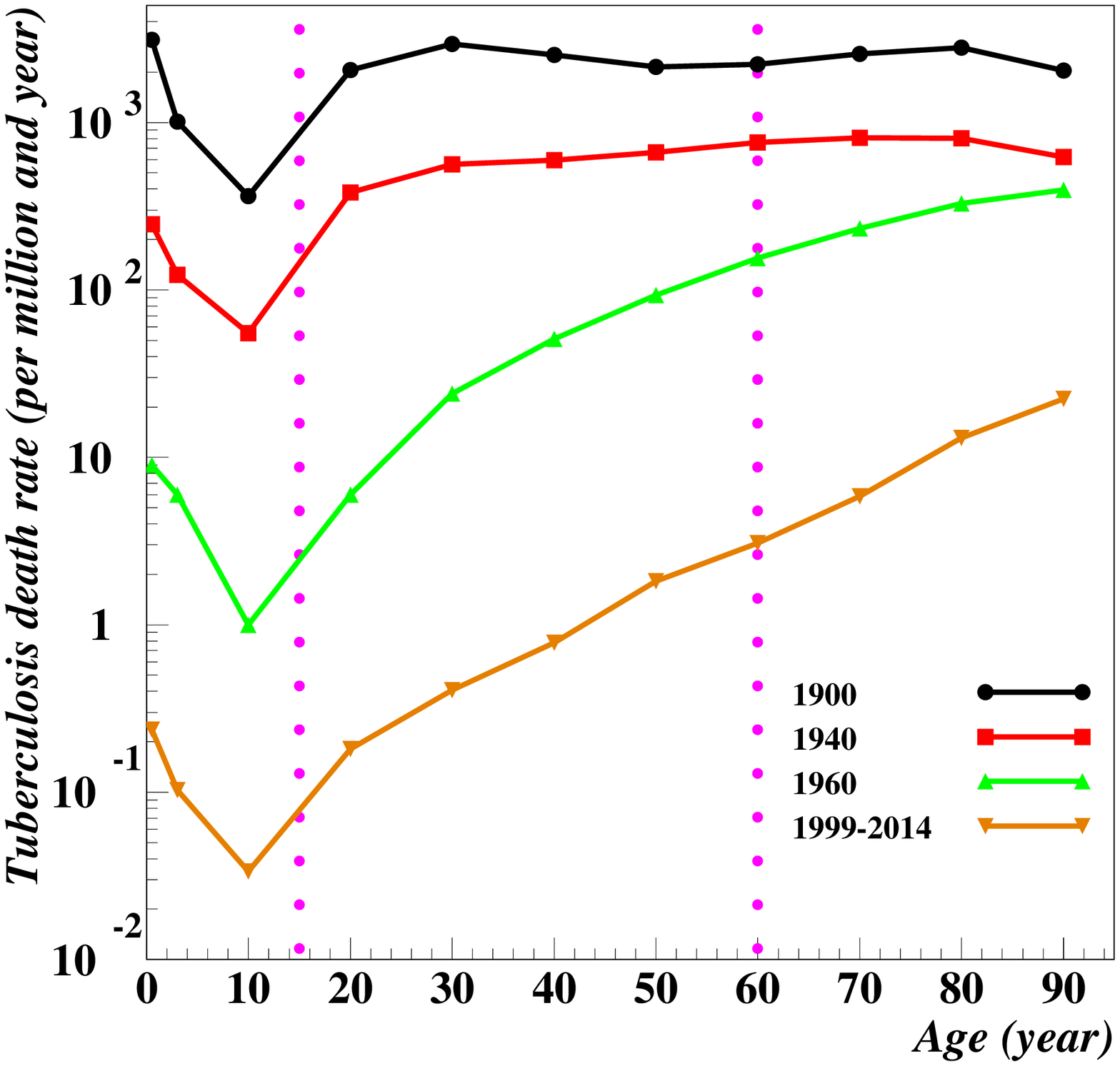}}
\qleg{Fig.\qhu 3\qhv TB 
death rate in the United States.}
{From 1940 to the end of the 20th century
the old-age component becomes more and more predominant.}
{Sources: Linder et al. (1947, p. 248-254);
Grove et al. (1968, p. 378-469); Wonder database of the
Center for Diseases control'' (CDC).}
\end{figure}

\begin{table}[htb]

\small

\centerline{\bf Table 1a\quad TB death ratios $ f/m $ and 
m/f in the age groups (0,15) and (15,60) in the US}

\vskip 5mm
\hrule
\vskip 0.7mm
\hrule
\vskip 2mm

$$ \matrix{
  \hbox{Year}  & \hbox{female/male} & \hbox{male/female} \cr
\qtb
   & I_1=(0,15) & I_2=(15,60) \cr
\noalign{\hrule}
\qth
 1910 & 1.2 & 1.33 \cr
 1931 & 1.1 & 1.27 \cr
 1943 & 1.2 & 1.71 \cr
 1950 & 1.2 & 2.16 \cr
 1954 & 1.1 & 2.41 \cr
\qtb
1999-2015 & - & 2.64 \cr 
\noalign{\hrule}
} $$
\vskip 1.5mm
\small
Notes: The symbol $ - $ means that the ratio is not well
defined because the numbers of deaths are too small.
It can be seen that for $ I_1 $ the ratio female/male
remains stable around a value of 1.1. In contrast
after 1931,
as the deaths in the sub-interval (35,60) of $ I_2 $ become
predominant, the ratio male/female increases from 1.3 to 2.6.
\qL
{\it Sources: Vital Statistics of the United States, Wonder
database of the ``Centers for Diseases Control'' (CDC).}
\vskip 2mm
\hrule
\vskip 0.7mm
\hrule
\end{table}
%

%
\begin{table}[htb]

\small

\centerline{\bf Table 1b\quad TB death ratios $ f/m $ and
$ m/f $ in the age groups (0,15), (15,60) in various countries}

\vskip 5mm
\hrule
\vskip 0.7mm
\hrule
\vskip 2mm

$$ \matrix{
  \hbox{Country}  \hfill & \hbox{female/male} & \hbox{male/female} \cr
\qtb
   & I_1=(0,15) & I_2=(15,60) \cr
\noalign{\hrule}
\qth
\hbox{China} \hfill & 1.0  & 1.60 \cr
\hbox{India} \hfill  & 1.2 & 1.50 \cr
\hbox{Philippines} \hfill& 0.91 & 1.64 \cr
\hbox{South Africa} \hfill & 0.84 & \color{magenta} 3.51 \cr
\qtb
\hbox{Thailand} \hfill & 1.0  & 1.67 \cr
\noalign{\hrule}
} $$
\vskip 1.5mm
\small
Notes: The data are for 2008.
The value 3.51 for $ I_2 $ in South Africa
is an outlier not only with respect
to the other countries but also with respect to the whole
range 1.3-2.6 displayed in Table 1a. Therefore, it is likely
that this figure is not correct; probably female deaths were
under-estimated by a factor of 2.
\qL
{\it Source: World Health Organization 2011: Mortality and burden of
disease estimates for WHO member states in 2008.}
\vskip 2mm
\hrule
\vskip 0.7mm
\hrule
\end{table}
\qpar

Fig. 3 shows that the parts of the curves in the age-interval
$ (0,15) $ keep the same structure. On the contrary,  
in the age-interval $ (15,60) $ 
the old age component becomes more and more predominant
especially in the decades after 1940. 
If one remembers that, as shown in Fig. 1b, $ f/m>1 $
in the age-interval $ (15,30) $ but $ f/m<1 $ in the
age-interval $ (35,60) $, it becomes cleat that
over the interval $ (15,60) $ the ratio $ m/f $
will become larger as old-age deaths become
predominant. This is indeed what appears in table 1a.
\qpar

With the exception of South Africa,
the figures by country given in Table 1b are consistent with the
longitudinal data for the United States given in Table 1a.
The case of South Africa shows that, most likely, 
female deaths were
under-reported by a factor of 2. With a $ h/f $ ratio
of 1.16, Iran is another outlier (this time the male deaths would
seem to have been under-reported); however in this case the
fact that only one decimal digit (male deaths=0.6, female
deaths=0.7) is given in the WHO report makes the conclusion
somewhat uncertain. 
\qpar
In other words the $ f/m $ effect reported in this
paper can be used to probe the reliability of the data
published by national statistical agencies. As a matter of fact,
according to the accompanying explanations, the data 
published in the WTO report are not exactly identical to
the figures provided by member states but have been revised by the WHO
to ensure ``cross national comparability''. This makes the discrepancy
observed in Table 1b even more surprising.

\qI{Conclusion}

The female/male death rate ratio for TB in all industrialized
countries for
which appropriate data during the late 19th century through to the
mid-20th century
are available is greater than unity for the age range 10-35
whereas for all other diseases
the ratio is less than unity. From our analysis we can deduce that in some
developing countries (we pointed out the case of South Africa)
deaths from TB are being under-reported by
factors as large as two%
\qfoot{It has been suggested that such under-reporting may be the
  result of stigma faced by women with TB in developing countries.}%
.
Once detailed data for developing countries become available we
shall be able to
test the veracity of this prediction. 

\vskip 5mm

{\bf Acknowledgments}\quad We wish to express our sincere
thanks to Ms. Justine Pooley of the ``Office of National
Statistics'' for introducing us to detailed
British mortality data.

\vskip 5mm

{\bf \large References}

\qparr
Anderson (M.) 1990: The social implications of demographic change, in:
Thompson (F.M.L.) editor: The Cambridge social history of Britain, 1750-1950.
Vol. 2: People
and their environment. Cambridge University Press, Cambridge.

\qparr
Berrut (S.), Pouillard (V.), Richmond (P.), Roehner (B.M.) 2016:
Deciphering infant mortality. 
Physica A 463, 400-426.

\qparr
Bronikowski (A.M.),
Altmann (J.),
Brockman (D.K.),
Cords (M.),
Fedigan (L.M.), 
Pusey (A.) 
Stoinski (T.),
Morris (W.F.), 
Strier (K.B.),
Alberts (S.C.)
2011:
Aging in the natural world: comparative data reveal similar mortality
patterns across primates.
Science 331,6022,1325–1328.

\qparr
Bunle (H.) 1954: Le mouvement naturel de la population
dans le monde de 1906 \`a 1936. [Vital rates from 1906 to 1936
in many countries in the world.]
Les \'Editions de l'Institut d'\'Etudes D\'emographiques, Paris.

\qparr
Clutton-Brock (T.H.), Isvaran (K.) 2007: 
Sex differences in ageing in natural populations of vertebrates
Proceedings of the Royal Society B: Biological Sciences 274, 1629, 
3097–3104.

\qparr
Garenne (M.), Leroy (O.), Beau (J.P.), Sene (I), Whittle (H.),
Sow (A.R.) 1991: Efficacy, safety and immunogenecity of two
high titer measles vaccines. A study in  Niakhar, Senegal, Final
Report, ORSTOM (Office de la Recherche Scientifique et Technique 
Outre-Mer), Population et Sant\'e, Dakar, S\'en\'egal. \qL
[The ORSTOM has become the IRD (Institut de Recherche pour le
D\'eveloppement) in 1998.

\qparr
Garenne (M.) 1994: Sex differences in measles mortality:
a world review.
International Journal of Epidemiology 23,3,632-642.

\qparr
Grove (R.D.), Hetzel (A.M.) 1968:  Vital statistics rates in the
United States 1940-1960. US Government Printing Office,
Washington DC.

\qparr
Hinde (A.) 2011: Sex differentials in mortality in nineteenth-century
England and Wales. 
Paper presented at the ``Economic History Society Conference'', Cambridge,
1-3 April 2011.

\qparr
Janssens (A.) 2017: A natural female disadvantage? Maternal
mortality and the role of nutrition related causes of death
in the Netherlands, 1875--1899. 
Paper presented at the 28th
IUSSP (``International Union for the Scientific Study of
Population''), Cape Town 29 October--3 November 2017.

\qparr
Linder (F.E.), Grove (R.D.) 1947: Vital statistics rates in the
United States 1900-1940. US Government Printing Office,
Washington DC.


\qparr
Preston (S.H.) 1976: Mortality in national populations.
Academic Press, New York.

\qparr
Thomas (T.Y.), Rajagopalan (S.) 2001: 
Tuberculosis and aging: a global health problem.
Clinical Infectious Diseases 33,7,1034-1039.

\end{document}